\documentclass{ws-ijbc}
\usepackage{ws-rotating}     % used only when sideways tables/figures are used
\usepackage{graphicx}
\usepackage{epstopdf}
\usepackage{multicol}

\graphicspath{{figures/}}

\begin{document}

%\catchline{}{}{}{}{} % Publisher's Area please ignore

\markboth{P. S. Skardal}{}

\title{Stability Diagram, Hysteresis, and Critical Time Delay and Frequency for the Kuramoto Model with Heterogeneous Interaction Delays}

\author{Per Sebastian Skardal}

\address{Department of Mathematics, Trinity College, 300 Summit St.\\
Hartford, Connecticut, 06106, USA\\
persebastian.skardal@trincoll.edu}

\maketitle

\begin{history}
\received{(to be inserted by publisher)}
\end{history}

\begin{abstract}
We investigate the dynamics of large, globally-coupled systems of Kuramoto oscillators with heterogeneous interaction delays. For the case of exponentially distributed time delays we derive the full stability diagram that describes the bifurcations in the system. Of particular interest is the onset of hysteresis where both the incoherent and partially synchronized states are stable for a range of coupling strengths -- this occurs at a codimension-two point at the intersection between a Hopf bifucration and saddle-node bifurcation of cycles. By studying this codimension-two point we find the full set of characteristic time delays and natural frequencies where bistability exists and identify the critical time delay and critical natural frequency below which bistability does not exist. Finally, we examine the dynamics of the more general system where time delays are drawn from a Gamma distribution, finding that more homogeneous time delay distributions tend to both promote the onset of synchronization and inhibit the presence of hysteresis.
\end{abstract}

\keywords{Coupled Oscillators, Synchronization, Time Delay, Hysteresis.}

\begin{multicols}{2}
\section{Introduction}\label{sec:01}

The dynamics of large systems of coupled oscillators, in particular their synchronization properties, represents an important area of study in nonlinear dynamics due to their utility in modeling a wide variety of natural and engineered phenomena~\cite{Winfree,Pikovsky}. Examples of synchronization of large ensembles of oscillatory units include rhythmic flashing of fireflies~\cite{Buck1988QRB}, circadian rhythms~\cite{Strogatz1987JMB}, cardiac pacemakers~\cite{Glass}, and power grids~\cite{Rohden2012PRL}. The Kuramoto model, which was designed as an analytically tractable alternative to Winfree's seminal model~\cite{Winfree1967JTB}, is a particularly important example where oscillators' states are each characterized by a single phase angle, each oscillator has its own natural frequency, and oscillators are globally coupled through a sinusoidal coupling function~\cite{Kuramoto}. Winfree's original model was designed to model a generic variety of biological rhythms, but the Kuramoto model has found even more applications, including additional biological systems, e.g., brain dynamics~\cite{Hoppensteadt1999PRL,Kirst2016NatComms}, as well as engineered systems, e.g., Josephson junctions~\cite{Wiesenfeld1998PRE} and power grids~\cite{Dorfler2013PNAS,Skardal2015SciAdv}. Since its introduction, researchers in the nonlinear dynamics community have thoroughly studied the dynamics of further extensions of the Kuramoto model, for example external noise~\cite{Sakaguchi1988PTP,Strogatz1991JSP} and bimodal frequency distributions~\cite{Kuramoto,Crawford1994JSP}, in order to better understand the mechanisms underlying emergence of collective phenomena. 

One important extension of the Kuramoto model involves time delays in the interactions between oscillators. The equations of motion for the time delayed Kuramoto model are given by
\begin{align}
\dot{\theta}_n = \omega_n +\frac{K}{N}\sum_{m=1}^{N}\sin[\theta_m(t-\tau_{nm})-\theta_n(t)],\label{eq:01}
\end{align}
where $\theta_n$ represents the phase angle of oscillator $n$ with $n=1,\dots,N$, $\omega_n$ is the natural frequency of oscillator $n$, which we assume is drawn from a distribution $g(\omega)$, $K\ge0$ is the global coupling strength, and $\tau_{nm}\ge0$ represents the interaction delay between oscillators $n$ and $m$, which we assume are independent and identically drawn (iid) from a distribution $h(\tau)$. The degree of synchronization is measured by the magnitude of the complex order parameter defined as
\begin{align}
z = re^{i\psi} = \frac{1}{N}\sum_{m=1}^Ne^{i\theta_m},\label{eq:02}
\end{align}
where the amplitudes $r\approx0$ and $r\approx1$ correspond to incoherent and partially synchronized states, respectively. Early investigations into the dynamics of Eq.~(\ref{eq:01}) focused on the case of a uniform time delay, i.e., $\tau_{nm}=\tau$ for all $n,m$, and observed the emergence of hysteresis, i.e., regions of multistability between incoherence and partial synchronization in parameter space, a dynamical property that is absent in the classical Kuramoto model without interaction delays~\cite{Kim1997PRL,Yeung1999PRL,Choi2000PRE,Montbrio2006PRE}. Further analytical progress characterizing the macroscopic dynamics such as bifurcation structure and stability diagram remained elusive, leaving several questions unanswered. In particular, what bifurcations mark the transitions between incoherence and partial synchronization? How do these bifurcations depend on the characteristic time delay and characteristic natural frequency? Given a positive characteristic time delay, is it always possible to observe hysteresis, or is there a critical characteristic time delay below which hysteresis not exist?

Recently, Ott and Antonsen~\cite{Ott2008Chaos,Ott2009Chaos} discovered a remarkable technique for reducing the dimensionality of large oscillator systems, thereby facilitating breakthroughs in the analytical descriptions in a wide variety of Kuramoto model extensions. Examples where this technique has yielded analytical progress includes systems with external forcing~\cite{Childs2008Chaos}, bimodal frequency distributions~\cite{Martens2009PRE,Pazo2009PRE}, community structure~\cite{Abrams2008PRL,Barreto2008PRE,Skardal2012PRE}, assortative and disassortative network structures~\cite{Restrepo2014EPL,Skardal2015PRE}, pulse-coupled oscillations~\cite{Pazo2014PRX,Luke2014FCN,Laing2014PRE}, positive and negative coupling strengths~\cite{Hong2011PRL}, and high-order coupling~\cite{Skardal2011PRE}. In Ref.~\cite{Lee2009PRL} Lee et al. demonstrated that this technique could be applied to the time delay case by allowing for delays to be heterogeneously distributed according to certain classes of a delay distribution $h(\tau)$. These results put on firmer ground the emergence of hysteresis and were subsequently used to explore the dynamics of further extensions of the model, including the spatially extended case~\cite{Laing2011PhysicaD,Lee2011Chaos} and adaptive coupling~\cite{Skardal2014PhysicaD}. However, a unified analysis of the original system with heterogeneous time delays remains lacking, specifically an analytical description of its stability diagram and bifurcation structure for general parameters, in particular different characteristic time delays. Moreover, while it is well known that interaction delays promote hysteresis, it remains unknown at precisely what point, e.g., at what value of the characteristic time delay and other system parameters, hysteresis first occurs. In this work we address these issues. First, we employ the ansatz of Ott and Antonsen to obtain the low dimensional system describing the macroscopic system dynamics. This reduced system allows for an analytically tractable stability analysis, which we use to study the bifurcations that occur in the system when time delays are drawn from an exponential distribution and derive the stability diagram. (We note that self-consistency approaches do not allow for such a stability analysis.) Next, we study in detail the properties of a codimension-two point that corresponds to the onset of hysteresis. The behavior of this codimension-two point reveals the critical time delay and critical natural frequency below which no hysteresis exists. Finally, we investigate the dynamics of the system when time delays are drawn from a more general family of distributions and compare these results to the exponential case.

The remainder of this article is organized as follows. In Sec.~\ref{sec:02} we briefly discuss some mathematical preliminaries. In Sec.~\ref{sec:03} we present our main results. First, we presenting a bifurcation analysis of the system and deriving the stability diagram for various parameter values. Next we investigate the behavior of a codimension-two point that illuminates the onset of hysteresis in the system, allowing us to determine the critical time delay and critical natural frequency. In Sec.~\ref{sec:04} we investigate the dynamics for more general cases of delay distributions. In Sec.~\ref{sec:05} we close with a discussion of our results.

\section{Preliminaries}\label{sec:02}

Here we briefly introduce a few important preliminaries. The analysis of the classical Kuramoto model is facilitated by the fact that the equations of motion can be simplified using the order parameter. To this end, we use the collection of time delayed order parameters~\cite{Lee2009PRL} defined as
\begin{align}
w_n=\rho_ne^{i\phi_n}=\frac{1}{N}\sum_{m=1}^Ne^{i\theta_m(t-\tau_{nm})},\label{eq:03}
\end{align}
which represents a similar mean-field order parameter as the order parameter in Eq.~(\ref{eq:02}), but delayed according to the time delays $\tau_{nm}$ ``felt'' by oscillator $n$. We will therefore refer to $z$ as the instantaneous order parameter and $w$ as the time delayed order parameter. Using this collection of time delayed order parameters, Eq.~(\ref{eq:01}) can be rewritten as
\begin{align}
\dot{\theta}_n&=\omega_n+ \frac{K}{2i}\left(w_ne^{-i\theta_n}-w_n^{*}e^{i\theta_n}\right)\nonumber\\&=\omega_n + K\rho_n\sin(\phi_n-\theta_n),\label{eq:04}
\end{align}
where $*$ represents the complex conjugate. Thus, from Eq.~(\ref{eq:04}) we see that the the role of the time-delays and time-delayed order parameters are crucial: each oscillator evolves according to its respective time delayed order parameter instead of the instantaneous order parameter given in Eq.~(\ref{eq:02}), as in the classical Kuramoto model.

Next we describe our choices of distributions for the time delays and natural frequencies. As noted above, we consider the case of heterogeneous time delays. Specifically, for the analysis presented in Sec.~\ref{sec:02} and \ref{sec:03} we assume that the distribution $h$ is exponential, i.e.,
\begin{align}
h(\tau) = \left\{\begin{array}{cc}e^{-\tau/T}/T&\text{ if }\tau\ge0,\\ 0& \text{ if }\tau<0,\end{array}\right.\label{eq:05}
\end{align}
where the parameter $T\ge0$ represents the characteristic time scale and mean for the interaction delays. In Ref.~\cite{Lee2009PRL} the authors considered the family of general Gamma distributions, of which the exponential is a special case. (We will investigate in more detail the case of general Gamma distributions in Sec.~\ref{sec:04}.) We will also restrict our attention to the case of a Lorentzian frequency distribution of the form
\begin{align}
g(\omega) = \frac{\Delta}{\pi\left[\Delta^2+(\omega-\omega_0)^2\right]},\label{eq:06}
\end{align}
where $\Delta>0$ represents the characteristic width of the distribution and $\omega_0$ represents the characteristic and mean natural frequency of the system. We note that in the classical Kuramoto model and several other extensions the analysis is facilitated by entering a rotating reference frame that effectively sets the characteristic natural frequency $\omega_0$ to zero. However, this cannot be done in this case due to the presence of the interaction delays in Eq.~(\ref{eq:01}) [and implicitly in Eq.~(\ref{eq:04})]. Therefore, as we will see below, both the characteristic time delay $T$ and the characteristic natural frequency $\omega_0$ will be key parameters in determining the system dynamics.

Finally, with these choices of delay and frequency distributions, the macroscopic system dynamics can be expressed using the dimensionality reduction discovered by Ott and Antonsen~\cite{Ott2008Chaos}. This reduction requires us to consider the continuum limit $N\to\infty$. Note first that in this limit, since time delays $\tau_{nm}$ are independent and identically drawn random variables, we have that $w_n=w$; that is, all $N$ time delayed order parameters collapse to the same value. Applying the dimensionality reduction then results in a closed-form system for $z$ and $w$:
\begin{align}
\dot{z} &= -\Delta z + i\omega_0 z + \frac{K}{2}\left(w-w^{*}z^2\right),\label{eq:07}\\
T\dot{w} &= z-w.\label{eq:08}
\end{align}
The derivation of Eqs.~(\ref{eq:07}) and (\ref{eq:08}) follows that presented in Refs.~\cite{Lee2009PRL,Laing2011PhysicaD,Lee2011Chaos,Skardal2014PhysicaD} and the details are presented in Appendix \ref{app:A}. Equation~(\ref{eq:07}) can be interpreted as the instantaneous order parameter evolving nonlinearly in reaction to the time delayed order parameter. Equation~(\ref{eq:08}) can be interpreted as incorporating the time delay, with the time delayed order parameter ``chasing'' the instantaneous order parameter at a time scale equal to the characteristic time delay $T$. Note that in the limit $T\to0^+$ we obtain $w=z$, which recovers the low dimensional description of the classical Kuramoto model when inserted back into Eq.~(\ref{eq:07})

\section{Bifurcation Analysis}\label{sec:03}

\subsection{Scaling}
We begin our analysis of the reduced dynamics in Eqs.~(\ref{eq:07}) and (\ref{eq:08}) with a rescaling to reduce the number of system parameters. In particular, we eliminate the parameter $\Delta$ by first defining
\begin{align}
\tilde{t} &= \Delta t,\label{eq:09}\\
\tilde{K} &= K/\Delta,\label{eq:10}\\
\tilde{\omega}_0 &= \omega_0/\Delta,\label{eq:11}\\
\tilde{T} &= \Delta T,\label{eq:12}
\end{align}
and then dividing Eqs.~(\ref{eq:07}) and (\ref{eq:08}) by $\Delta$, obtaining
\begin{align}
\dot{z} &= -z +i\omega_0 z + \frac{K}{2}\left(w-w^*z^2\right),\label{eq:13}\\
T\dot{w}&=z-w,\label{eq:14}
\end{align}
where the overdot now corresponds to differentiation with respect to rescaled time and we have dropped the $\sim$-notation for simplicity. This rescaling eliminates the parameter $\Delta$ by effectively fixing the width of the frequency distribution $g(\omega)$ to one with the remaining parameters appropriately rescaled. This rescaling essentially speeds up time, reduces the coupling strength and mean of the frequency distribution, and stretches the time delay distribution, all by a factor of $\Delta$.

\subsection{Steady-state dynamics and the $T=1$ case}
We now study the dynamics of the rescaled Eqs.~(\ref{eq:13}) and (\ref{eq:14}). Since both $z$ and $w$ are complex, an analytically tractable bifurcation requires us to convert to polar coordinates, i.e.,
\begin{align}
\dot{r} &= -\Delta r + \frac{K}{2}\rho(1-r^2)\cos(\phi-\psi),\label{eq:15}\\
\dot{\psi} &= \omega_0+\frac{K}{2}\rho\frac{1+r^2}{r}\sin(\phi-\psi),\label{eq:16}\\
T\dot{\rho} &=r\cos(\phi-\psi)-\rho,\label{eq:17}\\
T\dot{\phi} &=-\frac{r}{\rho}\sin(\phi-\psi).\label{eq:18}
\end{align}
We note that Eqs.~(\ref{eq:15})--(\ref{eq:18}) display a rotational invariance of the form $\psi\mapsto\psi+\delta$ and $\phi\mapsto\phi+\delta$, suggesting the existence of rotationally symmetric limit cycle solutions. These solutions can be found by searching for steady-state behavior with fixed amplitudes and angular velocities, i.e., setting $\dot{r}=\dot{\rho}=0$ and $\dot{\psi}=\dot{\phi}=\Omega$, where $\Omega$ represents the angular velocity. Inserting this into Eqs~(\ref{eq:17}) and (\ref{eq:18}) and using that $\cos^2x+\sin^2x=1$ yields
\begin{align}
\rho=\frac{r}{\sqrt{1+T^2\Omega^2}}.\label{eq:19}
\end{align}
This can in turn be inserted into Eqs.~(\ref{eq:15}) and (\ref{eq:16}) to yield a system of nonlinear equations that implicitly determines the steady-state values of $r$ and $\Omega$ for a limit cycle:
\begin{align}
r &=\frac{K}{2}\frac{r(1-r^2)}{1+T^2\Omega^2},\label{eq:20}\\
\Omega &=\omega_0 -\frac{K}{2}(1+r^2)\frac{T\Omega}{1+T^2\Omega^2}.\label{eq:21}
\end{align}
Inspecting Eqs.~(\ref{eq:20}) and (\ref{eq:21}), the incoherent state $r=0$ is always a trivial solution (in which case the angular velocity $\Omega$ has no physical meaning). Below we will present a stability analysis of the incoherent solution. We then search for nontrivial limit cycle solutions $r>0$ that correspond to partial synchronization. Moving forward we will consider cases where the mean natural frequency is positive, $\omega_0>0$, in which case it is also reasonable to search for solutions with a positive angular velocity, $\Omega>0$. (The analysis for $\omega_0<0$ runs similarly, in which case $\Omega<0$.)

\begin{figure*}[t]
\centering
\epsfig{file =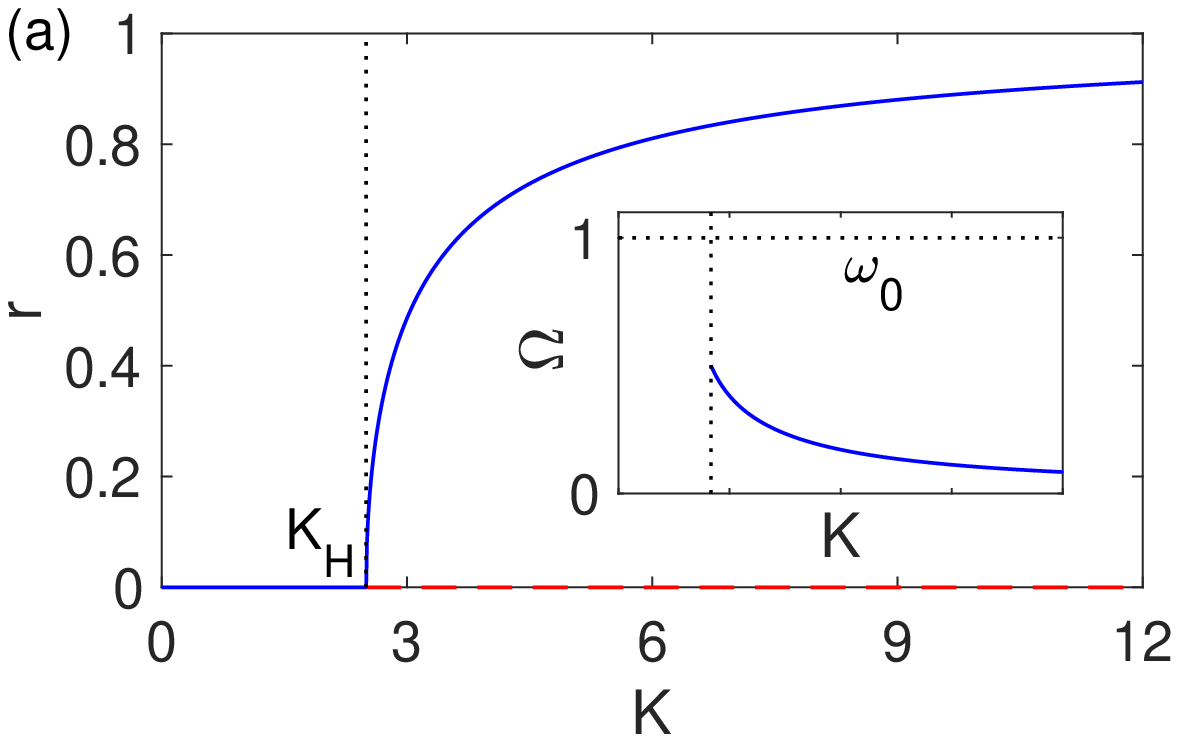, clip =,width=0.45\linewidth }
\epsfig{file =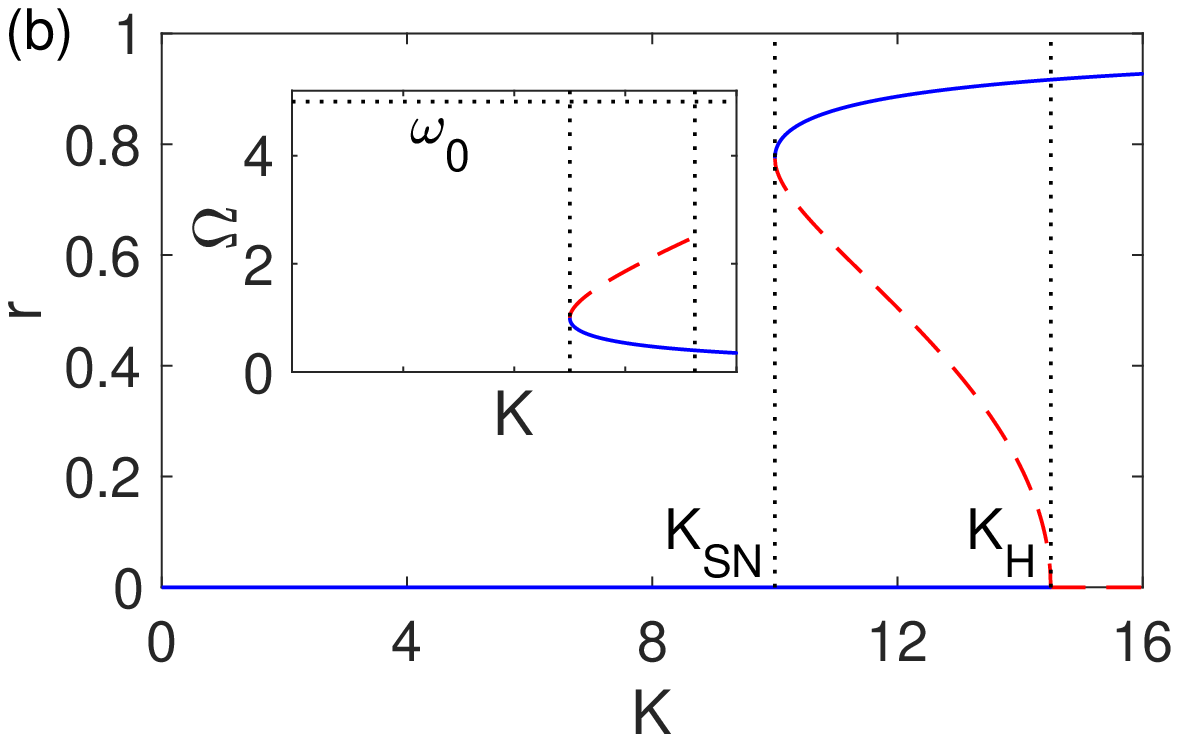, clip =,width=0.45\linewidth }
\caption{{\it Transitions to partial synchronization for $T=1$: supercritical and subcritical cases}. Steady-state solutions of $r$ vs $K$ with $T=1$ for (a) $\omega_0=1$ and (b) $5$, as examples of, respectively, a typical transition to partial synchronization through a supercritical Hopf bifurcation and via a hysteresis loop with a saddle node bifurcation of cycles and subcritical Hopf bifurcation. Stable and unstable branches are illustrated in solid blue and dashed red, respectively. Insets: steady-state solutions for the angular velocity $\Omega$ corresponding to partially synchronized solutions.}\label{fig1}
\end{figure*}

To shed light on the nature of nontrivial solutions, in particular the emergence of hysteresis, it is useful to briefly consider the case $T=1$, in which case analytical expressions for $r$ and $\Omega$ can be written down explicitly, as was done previously in Refs.~\cite{Laing2011PhysicaD,Lee2011Chaos,Skardal2014PhysicaD}. Setting $T=1$ and ignoring the incoherent state, we solve for $r$ in Eq.~(\ref{eq:20}), which we insert into Eq.~(\ref{eq:21}) and rearrange to find
\begin{align}
\Omega = \frac{K\mp\sqrt{K^2-4\omega_0^2}}{2\omega_0}.\label{eq:22}
\end{align}
Inserting this back into Eq.~(\ref{eq:20}) and neglecting the incoherent state $r=0$ yields
\begin{align}
r=\frac{\sqrt{\omega_0^2-K\pm\sqrt{K^2-4\omega_0^2}}}{\omega_0},\label{eq:23}
\end{align}
where we have also neglected the two negative solutions of $r$. We emphasize that the steady-state values of $r$ given in Eq.~(\ref{eq:23}) do not correspond to fixed points, but rather limit cycles with fixed amplitude $r$ and angular velocity $\Omega$. Inspecting Eq.~(\ref{eq:23}) more closely, the inner-most square root implies that a nontrivial solution for $r$ exists only if $K\ge2\omega_0$. Beyond this coupling strength, the solution in Eq.~(\ref{eq:23}) corresponding to the positive sign always exists, while the outer-most square root implies that the solution corresponding to the negative sign exists only if $K\le(\omega_0^2+4)/2$.

In Figs.~\ref{fig1}(a) and (b) we plot the solutions in Eq.~(\ref{eq:23}) for $\omega_0=1$ and $5$, respectively. Branches that are stable and unstable are plotted in solid blue and dashed red, respectively. For sufficiently small $\omega_0$, e.g., $\omega_0=1$, we see that the transition from incoherence to partial synchronization is typical in the sense that it is qualitatively similar to the transition in the classical Kuramoto model: a second-order phase transition from incoherence to partial synchronization owing to a supercritical Hopf bifurcation that occurs at a critical value we denote $K=K_{H}$. (We emphasize that this is in fact a Hopf bifurcation because the synchronized state is a limit cycle, not a fixed point.) However, for larger $\omega_0$, e.g., $\omega_0=5$, this transition folds over itself into a hysteresis loop as a saddle node bifurcation of cycles emerges at $K=K_{SN}<K_{H}$ and the Hopf bifurcation becomes subcritical. In this context, the term hysteresis refers to the multistability present for $K_{SN}\le K\le K_{H}$, where the initial conditions determine whether the system relaxes to the incoherent or synchronized state. Thus, a loop emerges, traversing the incoherent state and synchronized state, by repeatedly increasing the coupling strength beyond $K_H$ and decreasing it below $K_{SN}$. For the case of $T=1$ these critical bifurcation values can be characterized by the values for which the solutions in Eq.~(\ref{eq:23}) appear and annihilate. Specifically, the Hopf bifurcation occurs at $K_{H}=(\omega_0^2+4)/2$ and the saddle-node bifurcation of cycles occurs at $K_{SN}=2\omega_0$. Moreover, the saddle-node bifurcation of cycles emerges only if $\omega_0$ is large enough, i.e., larger than the intersection between $K_{H}$ and $K_{SN}$ which occurs at $\omega_0=2$. For the time being we forgo discussing the stability properties of these solutions, but will revisit this question below with a closer analysis of the Hopf bifurcation.

\subsection{Hopf and saddle node bifurcations}

We now seek to characterize the Hopf bifurcation and saddle node bifurcation of cycles illustrated for the $T=1$ case, but for general values of $T$. Specifically, we search for the critical coupling values $K_{H}$ and $K_{SN}$ at which these bifurcations occur. We begin with the Hopf bifurcation, which occurs when the partially synchronized solution $r>0$ collides with the incoherent solution $r=0$. To find this point we eliminate the incoherent solution from Eq.~(\ref{eq:20}), evaluate the limit $r\to0^+$, insert this into Eq.~(\ref{eq:21}), take the limit $r\to0^+$ again, and solve for $\Omega$ to obtain
\begin{align}\Omega = \frac{\omega_0}{1+T}.\label{eq:24}
\end{align}
Inserting Eq.~(\ref{eq:24}) back into Eq.~(\ref{eq:20}) (still in the $r\to0^+$ limit) and rearranging, we have that the Hopf bifurcation occurs at
\begin{align}
K_H = 2+\frac{2T^2\omega_0^2}{(1+T)^2}.\label{eq:25}
\end{align}
From Eq.~(\ref{eq:25}) we see that the onset of synchronization, which occurs at $K=K_H$, is earliest when $T=0$ and is delayed, i.e., occurs at larger coupling strengths, as the characteristic time delay $T$ and characteristic natural frequency $|\omega_0|$ are increased.

\begin{figure*}[t]
\centering
\epsfig{file =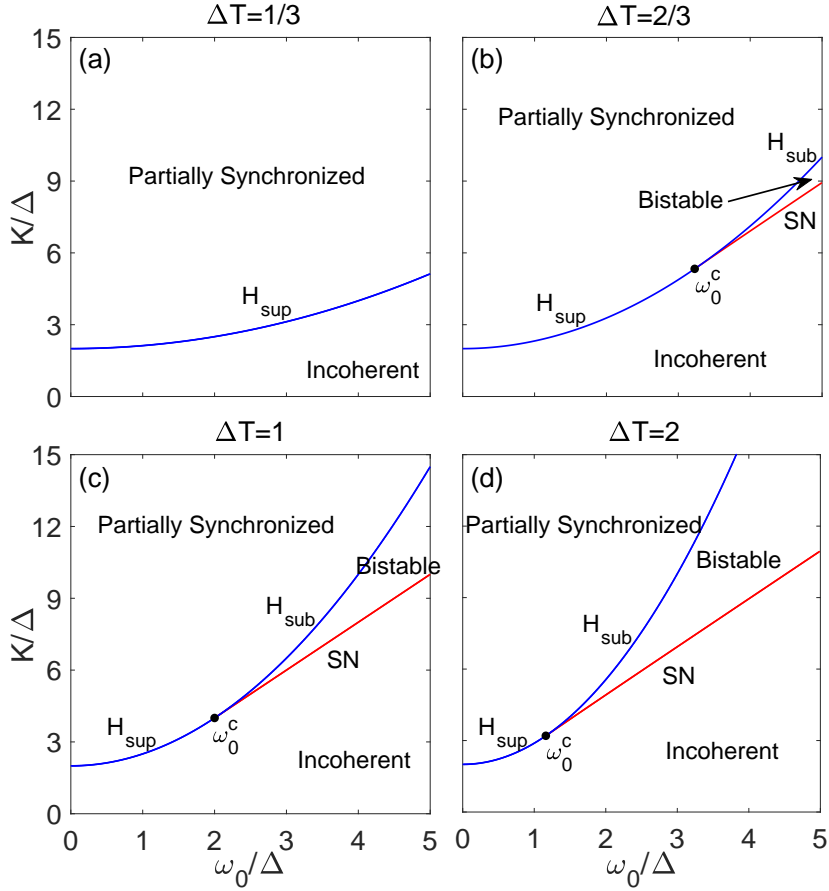, clip =,width=0.65\linewidth }
\caption{{\it Stability diagrams}. Stability diagrams in $(\omega_0/\Delta,K/\Delta)$ parameter space for (a) $\Delta T=1/3$, (b) $\Delta T=2/3$, (c) $\Delta T=1$, and (c) $\Delta T=2$. Blue curves denote Hopf bifurcations, labeled $H_{sup}$ and $H_{sub}$ for the supercritical and subcritical cases, respectively, and red curves denote saddle node bifurcations of cycles, labeled $SN$. These bifurcation curves partition the space into incoherent, partially synchronized, and bistable regions. The saddle node bifurcation collides with the Hopf bifurcations at a codimension two point labeled $\omega_0^c$.}\label{fig2}
\end{figure*}

Moving now to the saddle node bifurcation of cycles, we note that $K_{SN}$ coincides with a folding of both $r(K)$ and $\Omega(K)$ where $0=dK/dr=dK/d\Omega$. Choosing to work with $dK/d\Omega$, we eliminate the incoherent solution and solve for $r$ in Eq.~(\ref{eq:20}), which we insert into Eq.~(\ref{eq:21}) and rearrange to obtain
\begin{align}
K = \frac{(\omega_0-\Omega+T\Omega)(1+T^2\Omega^2)}{T\Omega}.\label{eq:26}
\end{align}
We then impose the constraint $dK/d\Omega=0$, which yields the expression (after multiplying by $T\Omega^2$ for convenience)
\begin{align}
2T^2(T-1)\Omega^3 + \omega_0T^2\Omega^2-\omega_0=0.\label{eq:27}
\end{align}
The bifurcation point $K_{SN}$ is then obtained by solving Eq.~(\ref{eq:27}) for $\Omega$ and inserting this back into Eq.~(\ref{eq:26}), however a few remarks are in order regarding this procedure. First, the left hand side of Eq.~(\ref{eq:26}) is cubic and thus it may have one or more roots that are difficult to express analytically; in practice we find it best to solve numerically. [Here we use Newton's method to find the roots of Eq.~(\ref{eq:27}).] Second, the nature of these solutions depends on the characteristic time delay $T$. Recall that $\omega_0>0$ and we search for a positive solution, $\Omega>0$. Note then that at $\Omega=0$ the left hand side of Eq.~(\ref{eq:27}) is negative, so that if $T\ge1$ only one such positive solution exists since the the derivative of the left hand side is positive for all $\Omega>0$. However, the $T<1$ case may admit an additional solution, or yield no solutions, depending on the values of $T$ and $\omega_0$. In particular, a positive local maximum will occur at $\Omega=\omega_0/3(1-T)$, and after inserting this back into Eq.~(\ref{eq:27}) we can see that a positive solution to Eq.~(\ref{eq:27}) will exist when $0<T<1$ only if $T\omega_0\ge3\sqrt{3}(1-T)$. Of these two roots we choose the smaller, which corresponds to the single positive root that exists for $T\ge1$. Finally, this solution is only valid if the corresponding solution for $r$ is real (and non-negative). To check this, we eliminate the incoherent solutions from Eq.~(\ref{eq:20}) and solve for $r$, yielding 
\begin{align}
r = \sqrt{1-\frac{2(1+T^2\Omega^2)}{K}}.\label{eq:28}
\end{align}
Thus, the saddle node bifurcation of cycles exists only if the solution to Eqs.~(\ref{eq:26}) and (\ref{eq:27}) satisfies 
\begin{align}
\frac{K}{2}\ge1+T^2\Omega^2.\label{eq:29}
\end{align}

\begin{figure*}[t]
\centering
\epsfig{file =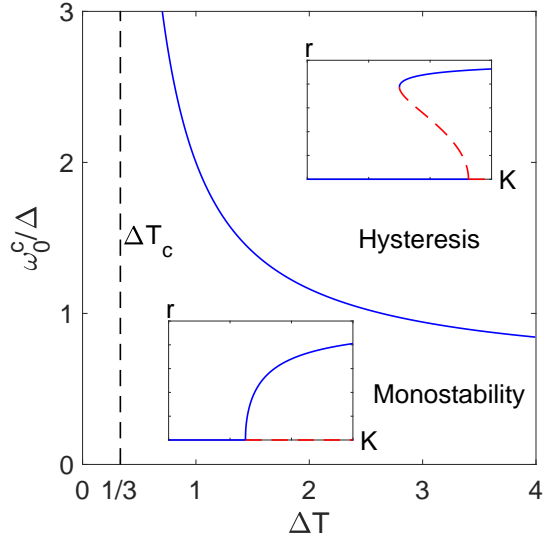, clip =,width=0.45\linewidth }
\caption{{\it Codimension two point and the emergence of hysteresis}. Behavior of the codimension two point $\omega_0^c/\Delta$ denoting the onset of hysteresis as the characteristic time delay $\Delta T$ varies. For $\omega_0<\omega_0^c$ and $\omega_0>\omega_0^c$ the transition to partial synchronization exhibits monostability and hysteresis, respectively, as illustrated by the insets. Below the critical time delay of $\Delta T_c=1/3$ no hysteresis is attainable for any values of $\omega_0$.}\label{fig3}
\end{figure*}

In Fig.~\ref{fig2} we illustrate the stability diagram in the parameter space $(\omega_0/\Delta,K/\Delta)$ for the system for a collection of representative characteristic time delays $T=1/3$, $2/3$, $1$, and $2$ in panels (a), (b), (c), and (d), respectively. (This and subsequent plots use the non-scaled parameters of the original system.) Hopf bifurcation curves are plotted blue and labeled $H_{sup}$ and $H_{sub}$ for the supercritical and subcritical cases, respectively, and the saddle node bifurcation of cycles curve is plotted in red and labeled $SN$. These bifurcation curves indicate the transitions between regions of parameter space corresponding to incoherence, partial synchronization, and bistability, where the bistable region is given by the wedge in between the subcritical Hopf and saddle node curves. Note first that for the case of $T=1/3$ no saddle node bifurcation of cycles exists in the visible window. In fact, no such bifurcation occurs for any value of $\omega_0$ when $T\le 1/3$ (which we shall see in the following section), so the transition to partial synchronization is supercritical as in Fig.~\ref{fig1} (a) for all $\omega_0$. For $T>1/3$ the saddle-node bifurcation of cycles persists for sufficiently large $\omega_0$, giving rise to a bistable region via a hysteretic transition to partial synchronization as in Fig.~\ref{fig1} (b). This bistability begins at the intersection of these curves, i.e., at a codimension-two point denoted with a filled black circle and labeled $\omega_0^c$. (We will discuss and study the behavior of this codimension-two point in detail in the following section.) We also observe that as $T$ increases the codimension-two point moves in with a smaller value of $\omega_0$, resulting in a larger bistable region. Physically, we can interpreted this behavior as an increase in either or both of the characteristic time delay $T$ and characteristic natural frequency $\omega_0$ promoting bistability in the system as it delays the onset of synchronization in the system by requiring a larger coupling strength $K$.

\subsection{Codimension two point and the emergence of hysteresis}

The results presented above highlight the importance of studying the properties of the codimension-two point corresponding to the onset of hysteresis. From the stabilty diagrams in Fig.~\ref{fig2}, the codimension-two point can be interpreted as the collision between the Hopf bifurcation and the saddle-node bifurcation of cycles. However, a more physical interpretation of this point can be better understood from Fig.~\ref{fig1}. In particular, for large enough $\omega_0$, the synchronization profile includes a hysteresis loop where both the incoherent and synchronized states are stable for $K_{SN}\le K\le K_{H}$. As $\omega_0$ is decreased, this range shrinks (which can be seen in Fig.~\ref{fig2}) until this interval vanishes, which occurs when the hysteresis loop in the synchronization profile ``unfolds''. The codimension-two point corresponds precisely to this ``unfolding''. We now seek to determine (i) for which values of $T$ the codimension-two point at $\omega_0^c$ exists, and (ii) if it does exist, then what value does it take? Recall that $\omega_0^c$ occurs at the intersection of the Hopf bifurcation and the saddle node bifurcation of cycles. Thus, $\omega_0^c$ is precisely the point at which multistability emerges: a hysteresis loop exists for all $\omega_0>\omega_0^c$, but not for $\omega_0<\omega_0^c$. Moreover, the existence of $\omega_0^c$ indicates the potential for hysteresis for a given time delay $T$. However, if $\omega_0^c$ does not exist for a given value of $T$, then multistability cannot be observed for any combination of the other system parameters.

To quantify the codimension two point $\omega_0^c$ we then search for the point where the saddle node bifurcation of cycles $K_{SN}$, which is a solution of Eqs.~(\ref{eq:26}) and (\ref{eq:27}), intersects with the Hopf bifurcation $K_{H}$ given in Eq.~(\ref{eq:25}). Despite the fact that $K_{SN}$ is given implicitly above, the point $\omega_0^c$ on this curve can be explicitly solved for. We begin by noting that the collision of the saddle node bifurcation with the Hopf bifurcation must occur in the limit $r\to0^+$. We thus eliminate the incoherent solution from Eq.~(\ref{eq:20}), solve for $K$, and insert this into the left hand side of Eq.~(\ref{eq:26}), yielding
\begin{align}
2=\frac{\omega_0+T\Omega-\Omega}{T\Omega}.\label{eq:30}
\end{align}
Equation~(\ref{eq:30}) is solved by $\Omega = \frac{\omega_0}{T+1}$, which can then be inserted into Eq.~(\ref{eq:27}) and rearranged to find that the codimension-two point given by
\begin{align}
\omega_0^c=\frac{(T+1)^{3/2}}{T\sqrt{3T-1}},\label{eq:31}
\end{align}
when it does exist.

The value of $\omega_0^c$ given in Eq.~(\ref{eq:31}) corresponds to the onset of hysteresis: when $\omega_0^c$ exists, then the system displays no bistability when $\omega_0\le\omega_0^c$, but when $\omega_0>\omega_0^c$ the transition from incoherence to partial stability goes develops a hysteresis loop. This is illustrated in Fig.~\ref{fig3}, where we plot $\omega_0^c/\Delta$ as a function of $\Delta T$ again using the non-scaled parameters of the original system. Below and above this curve we illustrate the transition to synchronization as supercritical and subcritical. Moreover, inspecting Eq.~(\ref{eq:31}) more closely, we see that the for $T\le1/3$ no real, finite codimension-two point exists due to the square root in the denominator. Therefore, bistability is not possible if the characteristic time delay is smaller than this critical value of $T_c=1/3$. This critical characteristic time delay is indicated with the dashed black curve In fact, in the limit $T\to T_c^+$ the codimension-two point $\omega_0^c$ approaches infinity, indicating that for characteristic time delays larger than, but close to $T_c$, extremely large characteristic natural frequencies are required to observe hysteresis. Moreover, in the limit of large $T$ we have that $\omega_0^c\to1/\sqrt{3}$, indicating that the mean natural frequency $\omega_0$ must be larger than this value to observe multistability, even for an arbitrarily large characteristic time delay $T$. Thus, we may conclude that if either $T<T_c=1/3$ or $\omega_0<1/\sqrt{3}$, then hysteresis can be ruled out -- only collectively large enough combinations of $T$ and $\omega_0$ result in bistability.

\subsection{Hopf bifurcation revisited: stability of the incoherent state}

In the last portion of our analysis, we revisit the Hopf bifurcation discussed above by studying the stability of the incoherent state $z=w=0$. Since the polar decompositions $z=re^{i\psi}$ and $w=\rho e^{i\phi}$ are singular at the incoherent state where $r,\rho=0$ (the phase angles $\psi$ and $\phi$ lose physical meaning at this point) it is more convenient to study the quantities $x$, $y$, $u$, and $v$ where $z=x+iy$ and $w=u+iv$. In this four dimensional state space, the linear stability of the incoherent state is governed by the eigenvalues of the Jacobian $DF$ evaluated at $x=y=u=v=0$, which is given by
\begin{align}
DF = \begin{bmatrix}
-1 & -\omega_0 & \frac{K}{2} & 0 \\
\omega_0 & -1 & 0 & \frac{K}{2} \\
\frac{1}{T} & 0 & -\frac{1}{T} & 0 \\
0 & \frac{1}{T} & 0 & -\frac{1}{T}
\end{bmatrix}.\label{eq:32}
\end{align}
The incoherent state is stable if all eigenvalues have negative real part, so we search for the critical value of $K$ where the eigenvalues of $DF$ with largest real part are purely imaginary. Moreover, our analysis above suggests that the change in stability occurs in the form of a Hopf bifurcation, where a pair of complex conjugate eigenvalues simultaneously crosses the imaginary axis. In general, the eigenvalues of $DF$ in Eq.~(\ref{eq:32}) are difficult to write down explicitly, so we study them in cases below.

We first consider, as we did above, the simplifying case of $T=1$, in which case the eigenvalues can in fact be written down explicitly:
\begin{align}
\lambda = -1\pm\sqrt{K-\omega_0^2\pm\sqrt{\omega_0^4-2K\omega_0^2}}\bigg/\sqrt{2},\label{eq:33}
\end{align}
where the four different combinations of $\pm$ correspond to the four different eigenvalues of $DF$. For real, non-negative values of $K$ and $\omega_0$, the real part of all four eigenvalues in Eq.~(\ref{eq:33}) is $\lambda_{\text{real}}=-1$ provided that $K\le\omega_0^2/2$. Beyond this value, the eigenvalues with largest real part are given by choosing the plus sign outside of both square roots in Eq.~(\ref{eq:33}). We then identify a Hopf bifurcation by searching for eigenvalues of the form $\lambda=\pm i\lambda_{\text{imag}}$, yielding the constraint
\begin{align}
\pm i\lambda_{\text{imag}}=-1+\sqrt{K-\omega_0^2\pm\sqrt{\omega_0^4-2K\omega_0^2}}\bigg/\sqrt{2}.\label{eq:34}
\end{align}
Moreover, at the Hopf Bifurcation we must have that $\lambda_{\text{imag}}=\Omega$, so that it is straight forward to check that the values $\Omega$ and $K$ given in Eqs.~(\ref{eq:24}) and (\ref{eq:25}), specifically
\begin{align}
\lambda_{\text{imag}}=\Omega_H&=\omega_0/(1+T)\nonumber\\
&=\omega_0/2,\label{eq:35}\\
K_H&=2+2T^2\omega_0^2/(1+T)^2\nonumber\\
&=(4+\omega_0^2)/2,
\end{align}
satisfy Eq.~(\ref{eq:34}).

For more general values of $T$ the analysis becomes more cumbersome. For starters, eigenvalues are more difficult to write down explicitly, so it is more convenient to work with the charcteristic polynomial of $DF$, which is given by
\begin{align}
\frac{[k-2(1+\lambda)(1+\lambda T)]^2+4\omega_0^2(1+\lambda T)^2}{4T^2}=0.\label{eq:37}
\end{align}
To check that a Hopf bifurcation occurs at the value $K=K_H$ as given in Eq.~(\ref{eq:25}), we search a pair of purely imaginary, conjugate eigenvalues $\lambda=\pm i\lambda_{\text{imag}}$ that solve Eq.~(\ref{eq:37}). Again, using that at the bifurcation $\lambda_{\text{imag}}=\Omega$, it can easily be checked that choosing $\lambda=\pm i\Omega$ from Eq.~(\ref{eq:24}) and $K=K_H$ from Eq.~(\ref{eq:25}) solves Eq.~(\ref{eq:37}). We have also checked numerically that the other two eigenvalues have negative real part, confirming the existence of a Hopf bifurcation at the value $K=K_H$ from Eq.~(\ref{eq:34}).

\section{General delay distributions}\label{sec:04}

We now consider the possibility of time delays drawn from a more general class of delay distributions $h(\tau)$. Following Ref.~\cite{Lee2009PRL} we consider the family of Gamma distributions given by
\begin{align}
h(\tau)=\left\{\begin{array}{cl} \frac{(n+1)^{(n+1)}\tau^n}{\Gamma(n+1)T^{n+1}}e^{-(n+1)\tau/T} & \text{if }\tau\ge0\\ 0 & \text{if }\tau<0,\end{array}\right.\label{eq:38}
\end{align}
which has mean $T$ and the parameter $n\ge0$ controls the standard deviation, which is given by $T/\sqrt{n+1}$. (Note that for $n=0$ we recover the exponential distribution given in Eq.~(\ref{eq:05}) for which the analysis above is valid.) Thus, larger and smaller values of $n$ can be interpreted as having, respectively, a more homogeneous or heterogeneous distribution of time delays. The choice of Gamma distribution for $h(\tau)$ is particularly convenient because it allows us to describe the dynamics of the time delayed order parameter $w$ using a differential equation similar to Eq.~(\ref{eq:08}). Restricting $n$ to be a non-negative integer, the result is an $(n+1)^{\text{th}}$-order ordinary differential equation of the form
\begin{align}
\left[\left(\frac{T}{n+1}\right)\frac{d}{dt}+1\right]^{n+1}w(t)=z(t).\label{eq:39}
\end{align}
For instance, the choice $n=1$ yields $(T^2/4)\ddot w+T\dot{w}+w=z$, the choice $n=2$ yields $(T^3/27)\dddot{w}+(T^2/3)\ddot{w}+T\dot{w}+w=z$, and so forth. The derivation of Eq.~(\ref{eq:39}) hinges on the fact that $h(\tau)$ in Eq.~(\ref{eq:38}) is a convenient form for the Laplace transform, and further details are described in Appendix~\ref{app:B}. Moreover, the rescaling of parameters detailed in Eqs.~(\ref{eq:09})-(\ref{eq:12}) remains valid for Eq.~(\ref{eq:07}) together with Eq.~(\ref{eq:39}).

Equation~(\ref{eq:39}) completes the low-dimensional description of the system dynamics with time delays drawn from a general Gamma distribution when paired with Eq.~(\ref{eq:07}). However, a direct analytical description of the dynamics for general $n$ remains problematic for two reasons. First, Eq.~(\ref{eq:39}) is an $(n+1)^{\text{th}}$-order differential equation. Converting to a system of first-order differential equations and including Eq.~(\ref{eq:07}) results in a system of $n+2$ complex variables, or $2n+4$ real-valued variables. Moreover, As illustrated in Ref.~\cite{Lee2009PRL}, the dynamics for larger values of $n$ admit more complicated behaviors -- in addition to the possibility of the coexistence of stable incoherent and partially synchronized states, there exists multiple and distinct partially synchronized states.

To gain some insight into the structure of bifurcations for $n>0$ we begin by converting Eq.~(\ref{eq:39}) to a system of first-order differential equations. For a given value of $n$, we introduce the new complex variables $w_1,\dots,w_{n+1}$ that are defied by $w_1=w$ and $w_i=\dot{w}_{i-1}$ for $i=2,\dots,(n+1)$. Next, using Eq.~(\ref{eq:39}) and the binomial theorem we obtain the system of equations
\begin{subequations}
\begin{align}
\dot{w}_1&=w_2,\\
\dot{w}_2&=w_3,\\
&~~\vdots\nonumber\\
\dot{w}_n&=w_{n+1},\\
\frac{T^{n+1}\dot{w}_{n+1}}{(n+1)^{n+1}}&=z-\sum_{j=0}^n\frac{T^j}{(n+1)^j}\binom{n+1}{j}\frac{d^j}{dt^j}w_j,
\end{align}\label{eq:40}
\end{subequations}

\noindent Combined with Eq.~(\ref{eq:07}), Eqs.~(\ref{eq:40}) then allows us to simulate the macroscopic system dynamics for any $n\ge0$ with the $n+2$ complex variables $z,w_1,w_2,\dots,w_{n+1}$.

\begin{figure*}[t]
\centering
\epsfig{file =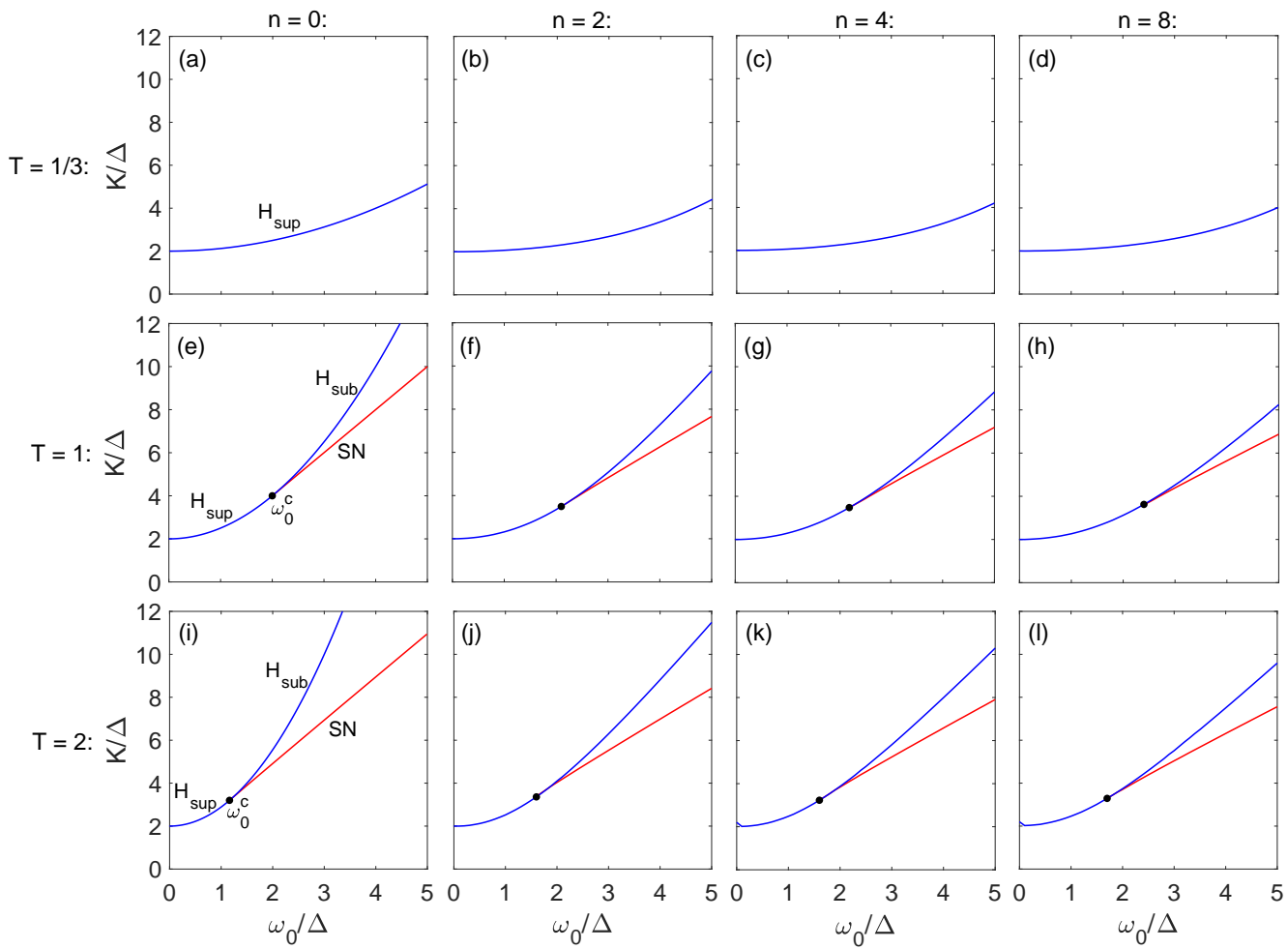, clip =,width=\linewidth }
\caption{{\it Stability diagrams for general time delay distributions}. Stability diagrams in $(\omega_0/\Delta,K/\Delta)$ parameter space for $\Delta T=1/3$ (top row), $\Delta T=1$ (middle row), and $\Delta T=2$ (bottom row). For each value of $\Delta T$ we show the results for $n=0$, $2$, $4$, and $8$, organized from left to right. Blue curves denote Hopf bifurcations and red curves denote saddle node bifurcations of cycles, with the codimension two point $\omega_0^c$ denoted with a black circle.}\label{fig4}
\end{figure*}

We now proceed numerically, first investigating the effect that varying $n$ has of the location of the Hopf bifurcation and saddle-node bifurcation of cycles. (Numerical simulation of Eqs.~(\ref{eq:40}) are obtained using Heun's method with a step size of $\Delta t=0.02$.) To identify the Hopf bifurcation we examine the stability of the incoherent state as follows. Starting at $K=0$, we perturb the incoherent state, $z=\delta z$, $w_i=\delta w_i$ (here we choose $|\delta z|,|\delta w_i|=10^{-3}$), and after a transient identify if the perturbation has grown or decayed. If the perturbation has decayed, we increase $K$ slightly and repeat until we verify that the incoherent state is unstable, allowing us to approximate $K_H$. To identify the saddle-node bifurcation of cycles, we start at some $K>K_H$ and starting with an initial condition $|z|=1$ and $|w_i|=1$, simulate through a transient to reach steady-state (thereby reaching the partially synchronized state with largest $z$). We then slowly decrease $K$ until the steady-state dynamics reach the incoherent state, identified by $|z|$ less than a small threshold value (here we used a threshold of $10^{-2}$). Then, if the value of $K$ we end up with is less than $K_H$ as computed previously, we identify $K_{SN}$ as this value, otherwise we deduce that no hysteresis exists, and therefore no $K_{SN}$ exists. For both $K_H$ and $K_{SN}$ we hone-in on a more precise value using a bisection algorithm.

\begin{figure*}[t]
\centering
\epsfig{file =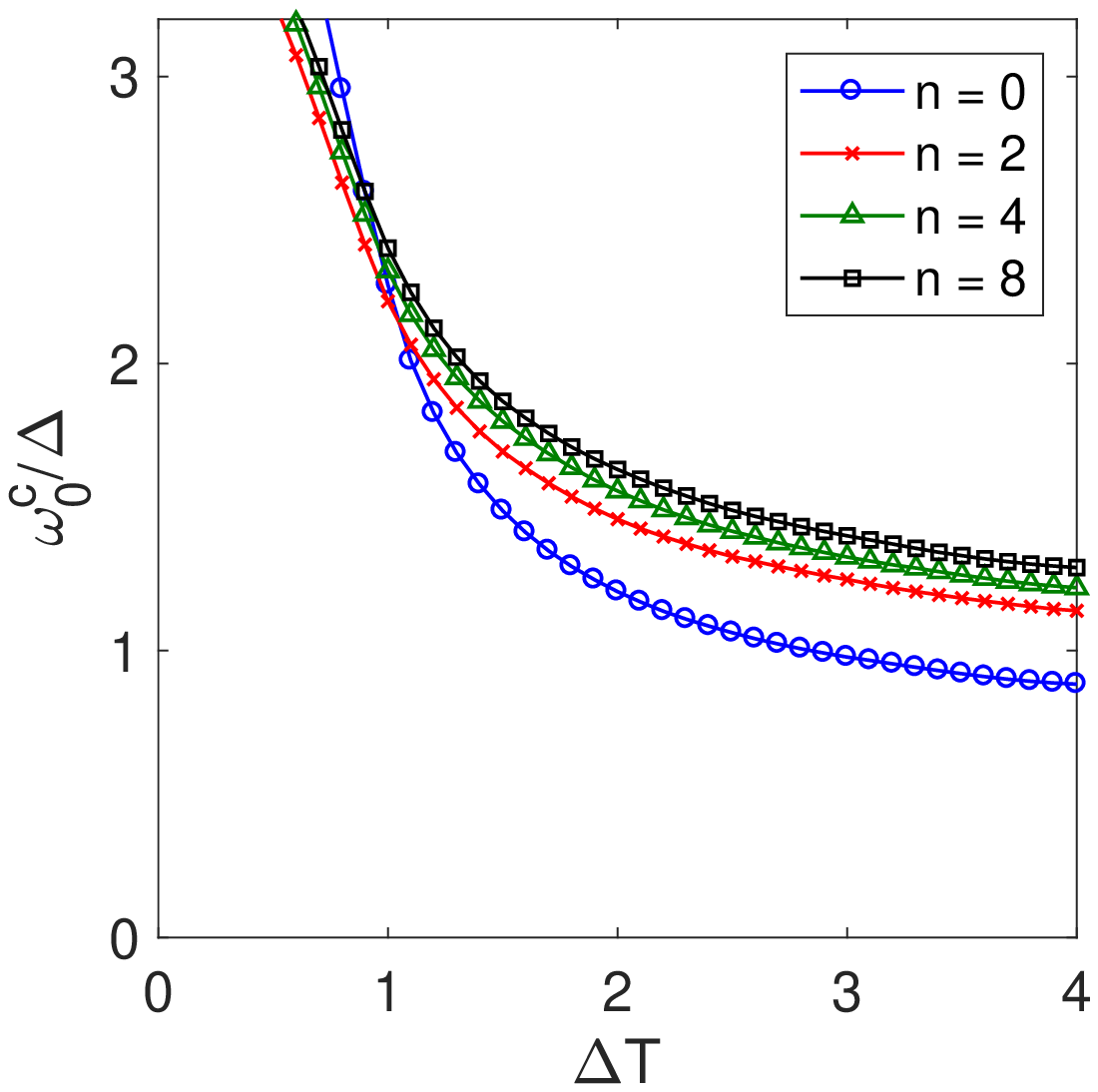, clip =,width=0.45\linewidth }
\caption{{\it Emergence of hysteresis for general time delay distributions}. Behavior of the codimension two point $\omega_0^c/\Delta$ denoting the onset of hysteresis as the characteristic time delay $\Delta T$ varies for delay distribution parameters $n=0$, $2$, $4$, and $8$. Hysteresis can be observed in the system only if $\omega_0>\omega_0^c$.}\label{fig5}
\end{figure*}

In Fig.~\ref{fig4} we plot a collection of 12 stability diagrams obtained by varying $T$ and $n$. The top row contains panels (a)--(d), corresponding to $T=1/3$ with $n=0$, $2$, $4$, and $8$, respectively, the middle row contains panels (e)--(h), corresponding to $T=1$ with $n=0$, $2$, $4$, and $8$, respectively,  and the bottom row contains panels (i)--(l), corresponding to $T=2$ with $n=0$, $2$, $4$, and $8$, respectively. Overall, the primary effect that increasing $n$ has on the dynamics is promoting synchronization. Specifically, as $n$ increases and the delay distribution $h(\tau)$ becomes more homogeneous, the Hopf bifurcation and saddle node bifurcation of cycles occur earlier, i.e., at small coupling strengths. For $T=1/3$ we found no trace of hysteresis, even for larger values of $\omega_0$ than those displayed in Fig.~\ref{fig4}. For $T=1$ and $2$ hysteresis persists although the codimension-two point $\omega_0^c$ increases, suggesting that larger $n$ delays the onset of hysteresis. However, this effect appears to be less pronounced than the change in the bifurcation curves themselves. We finally note that, because all results in Fig.~\ref{fig4} are obtained numerically, the left-most column, i.e., the $n=0$ case, serves as a numerical confirmation of the analytical results presented in Fig.~\ref{fig2}.

Next we investigate the codimension-two point characterizing the onset of hysteresis numerically using the following algorithm. For a given characteristic time delay $T$, we aim to identify the critical characteristic natural frequency $\omega_0^c$ for which the Hopf bifurcation first becomes subcritical. We begin with the incoherent state $z=0$ at $\omega_0=0$ and $K=0$. As we did above to identify the Hopf bifurcation, we increase $K$ until the incoherent state loses stability at $K_{H}$. Next we examine the dynamics of a state starting at $|z|=1$, $|w_i|=1$ at a slightly smaller coupling strength, $K_H-\Delta K$ (in our simulation we have used $\Delta K = 10^{-2}$). If the dynamics converge to the incoherent state (evaluated using the threshold of $r$ smaller than $10^{-2}$) then we determine that the bifurcation is supercritical and no hysteresis exists. In this case we increase $\omega_0$, reset $K=0$ and repeat the process above. If, on the other hand, the dynamics converge to a partially synchronized state, then we determine that the bifurcation is subercritical, indicating hysteresis, and let $\omega_0^c$ at the current value of $T$ be the current value of $\omega_0$. We repeat this process over a range of $T$ values to find a numerical description for the relationship between $\omega_0^c$ and $T$ for various values of the distribution parameter $n$.

In Fig.~\ref{fig5} we plot the results obtained numerically using our algorithm above describing the relationship between $\omega_0^c/\Delta$ and $\Delta T$ for $n=0$, $2$, $4$, and $8$ (blue circles, red crosses, green triangles, and black squares, respectively). Overall, we see that that for most characteristic time delays (roughly $T>1$) the critical natural frequency $\omega_0$ tends to increase as the distribution parameter $n$ increases. Recall that as $n$ increases, the distribution $h(\tau)$ becomes thinner while the mean remains at $T$. These results suggest then that, all else being equal, for sufficiently large $T$ the heterogeneity (that is, smaller $n$) in the delay distribution promotes hysteresis, while homogeneity (that is, larger $n$) in the delay distribution inhibits hysteresis. On the other hand, for smaller values of $T$ more homogeneous delay distributions appear to promote hysteresis over the more heterogeneous case of $n=0$. Moreover, this suggests that the critical characteristic time delay $T_c$ may be different for $n\ne0$, however using these numerical techniques it is difficult to gain insight into the behavior of $\omega_0^c$ for small or large $T$. Finally, because these results are computed numerically, the $n=0$ case serves as a numerical confirmation of the analytical results presented in Fig.~\ref{fig3}.

\section{Discussion}\label{sec:05}

In this paper we have studied the macroscopic system dynamics of the Kuramoto model of coupled oscillators with heterogeneous interaction delays. The effects of time delays on the Kuramoto model have been previously investigated~\cite{Kim1997PRL,Yeung1999PRL,Choi2000PRE,Montbrio2006PRE}, most notably giving rise to the possibility of multistability between incoherence and partial synchronization. Recently Lee et al.~\cite{Lee2009PRL} showed that the ansatz of Ott and Antonsen~\cite{Ott2008Chaos} could be applied to the time delayed case, but until this work no unified bifurcation analysis has been presented. Here we have presented such an analysis, allowing for the full description of the stability diagram for general parameter values which quantifies the transitions between incoherent, partially synchronized, and bistable states via a series of supercritical and subcritical Hopf bifurcations and a saddle node bifurcation of cycles. In particular, the transition from incoherence to partial synchronization occurs in one of two ways: (i) via a supercritical Hopf bifurcation in which case no bistability is observed, or (ii) via a subcritical Hopf bifurcation with a saddle node bifurcation of cycles in which case both the incoherent and partially synchronized states are  stable for a range of coupling strengths. We note that the nature of bistability studied here is reminiscent of explosive synchronization observed in networks of coupled oscillators whose natural frequencies are correlated with the network structure~\cite{GomezGardenes2011PRL,Peron2012PRE,Skardal2014PRE}.

In addition to the series of bifurcations that occur in the system, we also investigate the emergence of bistability. Specifically, this occurs at a codimension two point where the Hopf bifurcations collide with the saddle node bifurcation of cycles, which we express analytically. This codimension two point also reveals a critical characteristic time delay that delineates the possibility for bistability. In particular, if the characteristic time delay is less than this critical value, then bistability cannot be observed in the system, regardless of the choices of the other system parameters. If the characteristic time delay is larger than this critical value, then bistability can be observed, provided that the other system parameters are appropriately tuned. This suggests that the time delayed dynamics are only qualitatively different from the non-time delayed dynamics if the characteristic time delay in the system is sufficiently large. We have also used numerical techniques to investigate the dynamics that occur for more general delay distributions. First, we have observed that as the delay distribution (taken to be a Gamma distribution) becomes more homogeneous, synchronization is first promoted, as we can observed as both Hopf bifurcation and saddle node bifurcations of cycles occur at smaller coupling strengths. Second, for characteristic time delays that are not too small the emergence of hysteresis is delayed as the distribution becomes more homogeneous, as larger characteristic natural frequencies are required to observe hysteresis in the system. 

The results presented in this work shed light on the general behaviors of collective behavior as they depend on time delay. In many real-world scenarios it is realistic to incorporate time delays between interacting dynamical units, provided that either (i) a signal takes some finite time to travel from one unit to another or (ii) a given dynamical unit takes some finite time to interpret signals from others. Our analysis illustrates that the presence of time delays induces multistability, provided that the time delay is large enough. In a broader context these results raise the question of whether multistability or possibly other nonlinear effects can be induced on the collective behaviors of ensembles of dynamical units of other types, for instance in the contexts of consensus or spreading processes. Moreover, if time delays do induce new nonlinear effects in their collective behaviors, an important question is whether or not these nonlinear effects, like those found here, arise only for sufficiently large time delays larger than some critical value.

%\nonumsection{Acknowledgments} \noindent This part should come before References. Funding information may also be included here.

\appendix{Dimensionality Reduction}\label{app:A}

Here we detail the derivation of the low dimensional dynamics, i.e., Eqs.~(\ref{eq:07}) and (\ref{eq:08}) from the original system equations [Eqs.~(\ref{eq:01}) and (\ref{eq:04})] using the definitions of the instantaneous and time delayed order parameters [Eqs.~(\ref{eq:02}) and (\ref{eq:03})] and the chosen forms of the delay and natural frequency distributions [Eqs.~(\ref{eq:05}) and (\ref{eq:06})]. Following the technique presented in Refs.~\cite{Ott2008Chaos,Lee2009PRL}, we consider continuum limit, i.e., the limit of $N\to\infty$ oscillators. In this scenario, we may describe the macroscopic state of the system using the distribution function $f(\theta,\omega,t)$, such that $f(\theta,\omega,t)d\theta d\omega$ describes the fraction of oscillators with phases between $\theta$ and $\theta+d\theta$ and frequencies between $\omega$ and $\omega+d\omega$ at time $t$. In this continuum limit we may rewrite the instantaneous order parameter as
\begin{align}
z(t)=\int_{-\infty}^{\infty}\int_0^{2\pi}f(\theta,\omega,t)e^{i\theta(t)}d\theta d\omega.\label{eq:A01}
\end{align}
Since the time delays $\tau_{ij}$ are all drawn from the same distribution, we have that in this limit the time delayed order parameters are all equivalent, i.e., $w_i(t)=w(t)$ for all $i$, and can be written
\begin{align}
w(t)=\int_0^{\infty}z(t-\tau)h(\tau)d\tau.\label{eq:A02}
\end{align}
Moreover, the conservation of oscillators implies that the distribution function $f$ satisfies the following continuity equation:
\begin{align}
0&= \frac{\partial}{\partial t}f+ \frac{\partial}{\partial\theta}\left(f \dot{\theta}\right),\label{eq:A03}
\end{align}
where $\dot{\theta}=\omega + K\left(we^{-i\theta}-w^{*}e^{i\theta}\right)/(2i)$. Finally, because $f$ lives on the circle in the $\theta$ dimension it is natural to expand it into its Fourier series, which must be of the form
\begin{align}
f(\theta,\omega,t)=\frac{g(\omega)}{2\pi}\left\{1+\sum_{n=1}^\infty\left[f_n(\omega,t)e^{in\theta}+c.c.\right]\right\}.\label{eq:A04}
\end{align}

The dimensionality reduction discovered by Ott and Antonsen~\cite{Ott2008Chaos} consists of an ansatz for the sequence of the Fourier coefficients in Eq.~(\ref{eq:10}), specifically that they decay geometrically, i.e., $f_n(\omega,t)=a^n(\omega,t)$. Remarkably, inserting this ansatz into Eq.~(\ref{eq:A04}) then Eq.~(\ref{eq:A03}) reduces the partial differential equations in Eq.~(\ref{eq:A03}) to a single ordinary differential equation for $a$ of the form
\begin{align}
0=\frac{\partial a}{\partial t} + i\omega a + \frac{K}{2}\left(wa^2-w^{*}\right).\label{eq:A05}
\end{align}
Moreover the dynamics defined for the function $a(\omega,t)$ in Eq.~(\ref{eq:A05}) can be connected back to those of the instantaneous order parameter by inserting the Fourier series into Eq.~(\ref{eq:A01}), which yields
\begin{align}
z(t)=\int_{-\infty}^\infty g(\omega)a^{*}(\omega,t)d\omega.\label{eq:A06}
\end{align}
Recall now that we assumed a Lorentzian frequency distribution $g$, which can be rewritten
\begin{align}
g(\omega)=\frac{1}{2\pi i}\left(\frac{1}{\omega-\omega_0-i\Delta}-\frac{1}{\omega-\omega_0+i\Delta}\right).\label{eq:A07}
\end{align}
Using Eq.~(\ref{eq:A07}), the integral in Eq.~(\ref{eq:A06}) can be evaluated by closing the contour in the bottom-half $\omega$ complex plane containing the pole $\omega=\omega_0-i\Delta$, yielding $z(t)=a^{*}(\omega_0-i\Delta,t)$. (See Ref.~\cite{Ott2008Chaos}.) Finally, taking a complex conjugate of Eq.~(\ref{eq:A05}), evaluating at $\omega=\omega_0-i\Delta$, and rearranging yields
\begin{align}
\dot{z} = -\Delta z + i\omega_0 z + \frac{K}{2}\left(w-w^{*}z^2\right).\label{eq:A08}
\end{align}

In principle, Eq.~(\ref{eq:A08}) closes the dynamics of the system along with Eq.~(\ref{eq:A02}). However, for a bifurcation analysis it is convenient to convert the integral in Eq.~(\ref{eq:A02}) into a differential equation. In Ref.~\cite{Lee2009PRL} Lee et al. show that this is possible using a Laplace transform. In particular, Eq.~(\ref{eq:A02}) is a convolution, and therefore its Laplace transform is given by
\begin{align}
\hat{w}(s)=\hat{z}(s)\hat{h}(s),\label{eq:A09}
\end{align}
where $\verb!^!$ represents the Laplace transform. Since $h$ is exponential, i.e., $h(\tau)=e^{\tau/T}/T$ for $\tau\ge0$, we have that
\begin{align}
\hat{h}(s)=\frac{1}{1+Ts}\hskip2ex\to\hskip2ex (1+Ts)\hat{w}(s)=\hat{z}(s).\label{eq:A10}
\end{align}
We then convert back to the time domain and rearrange to obtain
\begin{align}
T\dot{w}=z-w,\label{eq:A11}
\end{align}
thus closing the dynamics of the system with Eqs.~(\ref{eq:A08}) and (\ref{eq:A11}), which are precisely Eqs.~(\ref{eq:07}) and (\ref{eq:08}) in the main text.

\appendix{Derivation of the Time Delay Equation for General Delay Distributions}\label{app:B}

Here we detail the derivation of Eq.~(\ref{eq:39}) in the main text, which describes the dynamics of the time delayed order parameter for the case of the general Gamma distribution $h(\tau)$ given in Eq.~(\ref{eq:38}). First, we note that in the limit of large system size, $N\to\infty$, Eq.~(\ref{eq:A02}) holds, now with Eq.~(\ref{eq:38}). Taking the Laplace Transform, we recover Eq.~(\ref{eq:A09}), but now with
\begin{align}
\hat{h}(s)=\left[\left(\frac{T}{n+1}\right)s+1\right]^{-(n+1)}\label{eq:B01}
\end{align}
Inserting Eq.~(\ref{eq:B01}) into Eq.~(\ref{eq:A09}) and rearranging yields
\begin{align}
\left[\left(\frac{T}{n+1}\right)s+1\right]^{n+1}\hat{w}(s)=\hat{z}(s),\label{eq:B02}
\end{align}
from which point we can convert back to the time domain, resulting in
\begin{align}
\left[\left(\frac{T}{n+1}\right)\frac{d}{dt}+1\right]^{n+1}w(t)=z(t),\label{eq:B03}
\end{align}
which is the desired result in the main text at Eq.~(\ref{eq:39}).

\end{multicols}
\end{document}